\documentclass[a4paper,fleqn,usenatbib]{mnras}
\usepackage[T1]{fontenc}
\DeclareRobustCommand{\VAN}[3]{#2}
\let\VANthebibliography\thebibliography
\def\thebibliography{\DeclareRobustCommand{\VAN}[3]{##3}\VANthebibliography}
\usepackage{graphicx}
\usepackage{amsmath}    
\usepackage{ae,aecompl}
\usepackage{amssymb}                    
\usepackage{lscape}
\usepackage{sidecap}
\usepackage{mathptmx}
\usepackage{txfonts}

\usepackage{color}                       
\usepackage{breakurl}                         

\newcommand{\oa}{    {\it Open Astron.}}
\newcommand{\lrsp}{    {\it Living Rev. Sol. Phys.}}
\chardef\us=`\_

\title[Joy's Law and Angular Momentum Transport] 
{Comparison Between Cycle-to-Cycle Variations in
the Coefficient of Joy's Law and Covariance of Rotation Residuals  
and Meridional Motions of Sunspot Groups}

\author[J. Javaraiah]{J. Javaraiah\thanks{E-mail: jajj55@yahoo.co.in; jj@iiap.res.in}
\thanks{Formerly worked in Indian Institute of Astrophysics, Bengaluru-560 034,
India}\\
 Bikasipura, BSK 5th Stage, Bengaluru-560 111, India}

\date{Accepted XXX. Received YYY; in original form ZZZ}

\pubyear{2015}

\begin{document}
 \overfullrule=0pt
\label{firstpage}
\pagerange{\pageref{firstpage}--\pageref{lastpage}}
\maketitle

\begin{abstract}
The tilts of bipolar magnetic regions are believed to be caused by the  action
of Coriolis force on rising magnetic flux tubes. Here we analysed the combined
 Greenwich  and Debrecen observatories sunspot-group data during the period
 1874\,--\,2017 and  the tilt angles of sunspot groups measured at Mt. Wilson
 Observatory  during the period 1917\,--\,1986 and Debrecen Observatory 
during the period 1994\,--\,2013. We find that there exists about 8-solar
 cycle (Gleissberg cycle) trend in the long-term variation of the slope 
of Joy's law (increase of tilt angle with latitude).  There exists a
 reasonably significant correlation between the slope/coefficient of Joy's law
and the slope  (namely, residual covariance) of the linear relationship
 between the rotation residuals and meridional motions of sunspot groups  
 in the northern hemisphere and also in the southern hemisphere during Solar
 Cycles 16\,--\,21. We also find that there exists a good correlation between
 north--south difference (asymmetry) in the coefficient of Joy's law and 
   that in the residual covariance.  We consider the residual
 covariance represents tentatively the coefficient of  angular
 momentum transport. These results
 suggest that there exists a relationship between the surface/subsurface
 poleward/equatorward angular momentum transport and the Joy's law. 
 There is a
 suggestion  of the strength of the Joy's law depends on the strength of the
 poleward angular momentum transport. 
\end{abstract}




\section{Introduction}

\cite{ward65}  analyzed
Greenwich sunspot-group data during the period 1935\,--\,1944 and found a
 significant correlation between the angular and meridional velocities of
 sunspot groups and interpreted it as meridional flows transfer angular 
momentum toward equator. Later many scientists studied this correlation by 
using various data and methods
 \citep[][and references therein]{sudar14,sudar17,jj21}.
Theoretical models predicted this correlation  as reflection of equatorial
angular momentum transport caused by Reynolds stresses near the surface
\citep[e.g.][]{gilman86}. 
Earlier, \citep{jj21}, we analysed the combined Greenwich and DPD sunspot-group
data during the period 1874\,--\,2017 and studied solar cycle-to-cycle
variation  in the slope  (namely, ``residual covariance") 
of the linear relationship between the rotation
residuals and meridional motions of sunspot groups  in the Sun's whole-sphere.
 In the present analysis we analysed the same sunspot-group data and
 determined the  residual covariance and its solar
 cycle-to-cycle modulations in northern and southern hemispheres. 

The line joining the leading and following
parts of an active region makes an angle to the parallel
of the local latitude. This angle  is referred to as tilt angle of
the active region.
Various dynamic
characteristics of active regions seem to be depend on the tilt angles of
active regions~\citep{howard91,howard96a}.
 The average tilt angle of active  regions increases with latitude
 \citep{hale19}. It is referred to as Joy's law and  has been studied
 extensively~\citep[for references see][]{gao23}. 
 Recently, \citep{jj23}, we analysed Mt. Wilson Observatory (MWOB)
 sunspot-group  data during the period 1917--\,1986 and
studied the solar cycle-to-cycle modulations in the coefficient of
Joy's law during Solar Cycles 15\,--\,21 in the  whole-sphere and northern
and southern hemispheres.

The tilts of bipolar magnetic regions are believed to be caused by the  action
of Coriolis force on rising magnetic flux tubes.  
 Coriolis force may be responsible for Joy's
law and hence the latter could  be related to the equatorward/poleward
angular momentum transport.     
 In the present analysis, in order to check
whether there exists a relationship between the cycle-to-cycle variations
in the Joy's law and  the poleward/equatorward angular
momentum transport, we determined correlation  between the cycle-to-cycle
 variations in
the coefficients of Joy's law and  residual covariance 
 during  Solar Cycles 15\,--\,21 determined from the 
 whole-sphere data and the northern and southern hemispheres' data separately.
 We consider the residual covariance as the 
 coefficient of angular momentum transport.

In the next section we describe the data and analysis. In Section~3
we describe the results and in Section~4 we present the conclusions
and briefly discuss them.

\section{Data Analysis}
\subsection{Determination of the "Residual Covariance" From 
Greenwich and DPD Sunspot-group Data}
Earlier,  \cite{jj21}, we analysed the  combined Greenwich Photoheliographic 
Results (GPR) and Debrecen Photoheliographic Data (DPD) of sunspot-group
 during 1874\,--\,2017.
The solar sidereal angular velocity $\omega$ (in degree day$^{-1}$) 
was calculated as 
$\omega(\theta) = \frac{L_i-L_{i -1}}{t_i - t_{i-1}} + 14^\circ.18$, 
and the meridional velocity
$v_{mer}$ (in degree day$^{-1}$) of the sunspot group  
was calculated as 
$v_{mer} (\theta) = \frac{\lambda_i-\lambda_{i-1}}{t_i - t_{i-1}}$,
where $L_{i -1}$  and $L_{i}$  are heliographic longitudes, and 
$\lambda_{i-1}$ and $\lambda_{i}$ are heliographic latitudes of a sunspot 
group measured at
times $t_{i-1}$ and $t_{i}$ during  the life time (disk passage) of the
sunspot group, $\theta = \lambda_{i-1}$, and
$14^\circ.18$ day$^{-1}$ is the Carrington rigid body
rotation rate. These daily values in each year during 1874\,--\,2017 were
 binned into different $5^\circ$ latitude intervals 
(0\,--\,$5^\circ$),
(5\,--\,$10^\circ$),\dots,(35\,--\,$40^\circ$)
of the northern and southern hemispheres,
 and calculated the mean values of $\langle \bar \omega \rangle (\lambda)$ and
 $\langle \bar v_{mer} \rangle (\lambda)$  in each latitude interval
  in which the data were available, 
  where $\lambda$ is the middle value of a latitude interval 
 (note that `$-$' indicates mean over a latitude interval  and 
`$\langle . \rangle$' indicates mean over time, a year).
 Using the  $\langle \bar \omega \rangle (\lambda)$ of all the 
latitude intervals in all 
the years during  the whole period 1874\,--\,2017 we 
 obtained the grand differential rotation law,
$\langle \bar \omega \rangle (\lambda)
   = (14.5 \pm 0.01)-(2.2 \pm 0.07) \sin^2{\lambda}$
 degree/day.
For each  latitude interval of a year we determined the residual rotation
 $\Delta \langle \bar \omega_{rot} \rangle (\lambda) =
 \langle \bar \omega\rangle (\lambda) - \xi (\lambda)$, where
 $\xi (\lambda)$ is the value of $\langle \bar \omega\rangle (\lambda)$ 
 deduced from  the differential rotation law. 
The $\Delta \langle \bar \omega\rangle (\lambda)$ 
degree/day was converted into  $\Delta \langle \bar v_{rot} \rangle (\lambda)$ 
meter/second and the values of  $\langle \bar v_{mer} \rangle (\lambda)$ was
 also  converted  to the unit meter/second. The values of 
 $\Delta \langle  \bar v_{rot} \rangle (\lambda)$ and 
$\langle \bar v_{mer} \rangle (\lambda)$
  of all the years of a solar cycle were fitted to the following 
 linear equation:  
\begin{equation}
\label{eq.1}
\langle \bar v_{mer} \rangle = D \Delta \langle \bar v_{rot} \rangle + C.
\end{equation}
 From this equation we have the following possibilities:
When $\langle \bar v_{rot} \rangle$ is a positive value,
a  positive/negative  $D$ mostly produces a  positive/negative
 $\langle \bar v_{mer} \rangle$ (poleward/equatorward motion).
Obviously,  it is  opposite  
when  $\langle \bar v_{rot} \rangle$ is a negative value.
  That is, the slope $D$  (namely, residual covariance) is a 
dimensionless quantity, its sign indicates the direction of angular
 momentum transport. Therefore, we consider that the slope $D$ 
 represents tentatively  the coefficient of poleward/equatorward 
angular momentum transport.
A negative/positive value of  $D$ indicates 
 equatorward/poleward angular momentum transport 
\citep[also see][]{sudar14,sudar17}. 

Note that the correlation between
 the two horizontal forces (the `covariance'), the rotational and meridional
 motions, may represent angular momentum transport toward equator by Reynolds
 stresses that is believed to be responsible for solar differential rotation.
Numerical simulations by \cite{brun04} suggest that the Maxwell tresses
 related to the magnetic fileds tend to oppose the Reynolds stresses
 causing poleward angular momentum transport. It may be also worth to 
note that  sunspots exhibit both equatoward and poleward meridional motions
\citep{ju06}.

  The data reduction was 
 carried out with  all the necessary precautions that were taken care  
in \cite{jj21}. In that paper  we have studied  
the solar cycle-to-cycle modulation in $D$ determined from the 
whole-sphere's sunspot-group data during Solar Cycles 12\,--\,24 
(note that in the case of Solar Cycle 24 the data were incomplete). 
Here   the northern and southern hemispheres' data are separately fitted 
to equation~\ref{eq.1}  and 
  determined the northern and southern hemispheres' values of $D$.
However, we find that there exists no significant correlation between the 
amplitude ($R_{\rm M}$) of a solar cycle  and the corresponding slope 
$D$ of the whole sphere as well as that of any  hemisphere. 

\subsection{Recovering Joy's law From MWOB Sunspot-group Data}
Recently, \cite{jj23},  we analysed the daily sunspot-group data measured 
in MWOB 
during the period 1917\,--\,1986. We binned the daily
  sunspot-group  tilt angle ($\gamma$)  data to the 
different $5^\circ$-latitude bins (absolute) 0\,--\,$5^\circ$, 
5\,--\,$10^\circ$,\dots,25\,--\,$30^\circ$, 30\,--\,$35^\circ$. 
First  the whole-sphere's data were analysed and then 
the northern and southern hemispheres' data 
were analysed separately. We determined the average values of tilt
angle ($\langle \bar \gamma \rangle$) in
each latitude bin during a solar cycle and they were fitted to the following 
linear equation 
(here `$\langle . \rangle$' indicates mean over a solar cycle): 
\begin{equation}
\langle \bar \gamma \rangle   = m |\lambda| + c,
\label{eq.2}
\end{equation}
where $|\lambda|$ is the absolute middle value of a latitude interval and 
 the slope $m$ represents the coefficient of Joy's law. A positive value 
of $m$ implies leading group closer to the equator than  following group.
 We studied 
solar cycle-to-cycle modulations in  $m$ determined from 
the whole-sphere data and  also that determined separately 
 from the northern and southern hemispheres' data  in 
each of Solar Cycles 15\,--\,21 (note that the data of 
first 4 years of  Solar Cycle 15  and  the last year of 
Solar Cycle 21 are missing). Only the southern hemisphere's 
  $m$ of a solar cycle was found to be reasonably well correlate to the 
amplitude ($R_{\rm M}$) of the solar cycle.

\subsection{Determination of Correlation Between $m$ and $D$}
Here we have used the values of  
$m$ determined in \cite{jj23} for all the three cases: whole-sphere,  
northern hemisphere, and southern hemisphere. 
We have used the values of  the whole sphere's $D$ that are taken from
 Table~2 of \cite{jj21}  and the corresponding hemispheres' values that 
are determined in the present analysis.
 We determined correlation between solar cycle-to-cycle 
variations in $m$ and $D$ during Solar Cycles 15\,--\,21 for 
 aforementioned all the three cases. 
Except in the case of equation~\ref{eq.1}, in each of the remaining linear
 regression analyses presented in this article
the uncertainties in both abscissa and ordinate are  taken care in
the calculation of the linear-least-square fit. For this we used
 the Interactive Digital Library (IDL) software {\textsf{FITEXY.PRO}},
 which  is downloaded from the website 
\textsf{idlastro.gsfc.nasa.gov/ftp/pro/math/}.
  In the case of equation~\ref{eq.1}, when we have used the uncertainties  in
$\Delta \langle  \bar v_{rot} \rangle$ and $\langle \bar v_{mer} \rangle$
the corresponding least-square best-fits were found not good.

 In \cite{jj23} 
we studied the variations in $m$ determined from both the tilt-angle data 
and the area-weighted tilt-angle data (the corresponding 
values were given in Tables~3\,--\,5 of that 
paper). However, here we have used only the values of $m$ that were
determined from the tilt-angle data. This is because no reasonable  
correlation between $m$ and $D$ was found when we have used 
the values of  $m$ derived from the area weighted tilt-angle data, 
 in all the three cases: whole sphere, northern hemisphere, and 
southern hemisphere. 

\subsection{Recovering Joy's law From DPD Sunspot-group Data}
Recently, \cite{gao23} has analysed 
the DPD sunspot-group tilt-angle data 
 during the period 1994\,--\,2013 and has found that the cyclical behaviors
of the tilt angles during Solar Cycle~23 are different from those of 
Solar Cycles~21 and 22.
Here we  also analysed the same data 
(downloaded from {\sf http://fenyi.solarobs.csfk.mta.\break hu/test/tiltangle/dpd/})
 and determined the values of $m$ 
(the slope of equation~\ref{eq.2})
 for Solar Cycle~21\,--\,23 for all the three cases: whole sphere, northern 
hemisphere, and southern hemisphere. For the detailed description of the 
DPD data can be found in~\citep{bara15,bara16,gyr17}.
 We analysed 
the data of  the tilt angles correspond to the  whole-spot areas of
 sunspot groups and also
 the tilt angles correspond to only the umbrae areas of
 sunspot groups.
The latter is similar to the  MWOB tilt-angle data.  We have used
 the tilt angles of sunspot groups whose leading and following parts were  
 separated by $> 2.5^\circ$. This is because the data correspond to the
 separations $\le 2.5^\circ$  contain large number of  unipolar sunspot
 groups~\citep{bara15}. We calculated the polarity separation ($\Delta s$)
 by using the  formula given by \cite{jiao21}: 
$\cos(\Delta S) = \sin(\lambda_f) \sin(\lambda_l) + 
\cos(\lambda_f) \cos(\lambda_l) \times \cos(\phi_f - \phi_l)$, 
where $\lambda_f$ and $\lambda_l$ are the latitudes of the following 
and leading polarities, respectively, and $\phi_f$ and $\phi_l$
are the corresponding longitudes.
 The sign convention of the DPD tilt angles are the same as those of 
MWOB, i.e. a positive value
of $m$ implies leading group closer to equator than  following group, in both 
the northern and southern hemispheres.
We have excluded the data correspond to the central meridian distance 
(CMD) $>60^\circ$.  In the case of DPD data we found a reasonable amount of 
the data also in the  latitude interval 35\,--\,$40^\circ$ and hence
 we have 
used this interval data also in the determination of the values of $m$ in all 
the three cases: whole sphere,  northern hemisphere, and southern hemisphere, 
 whereas in this latitude interval the MWOB data were found to be absent or
inadequate in all the three cases and hence they were not considered.
   
\begin{figure}
\centering
\includegraphics[width=8.5cm]{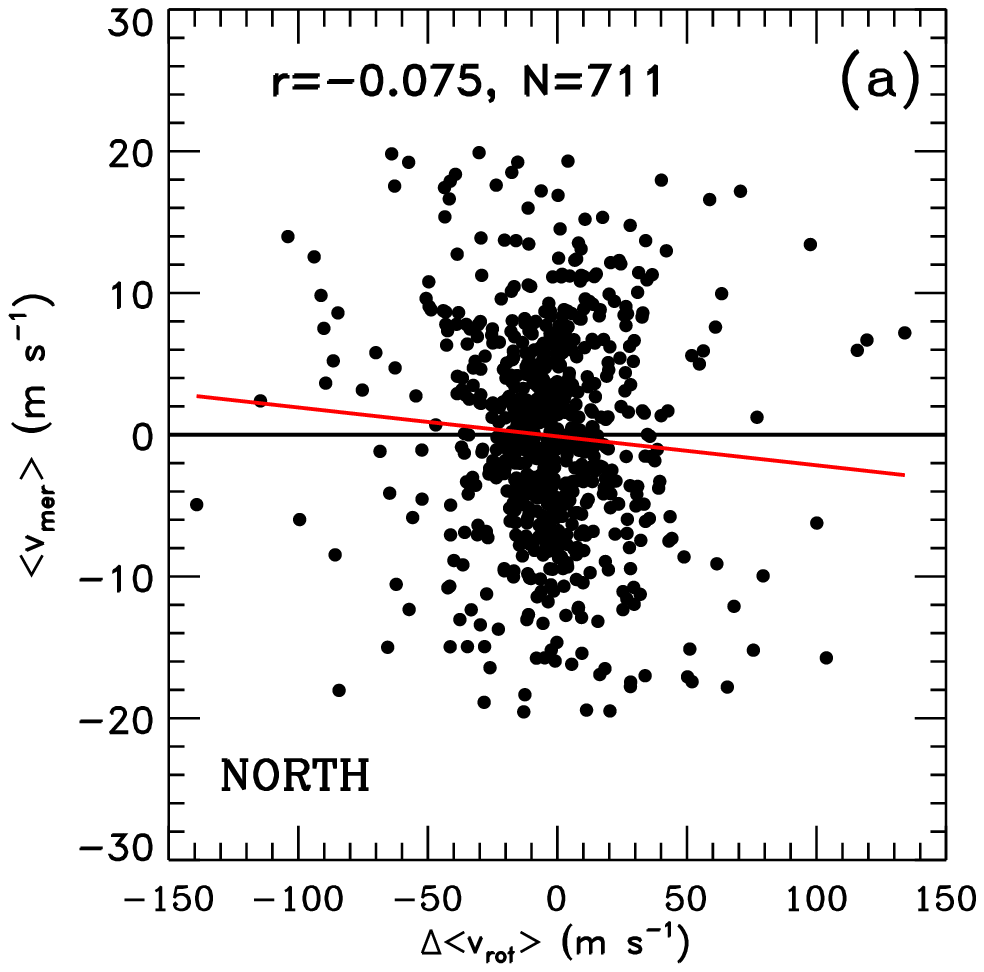}
\includegraphics[width=8.5cm]{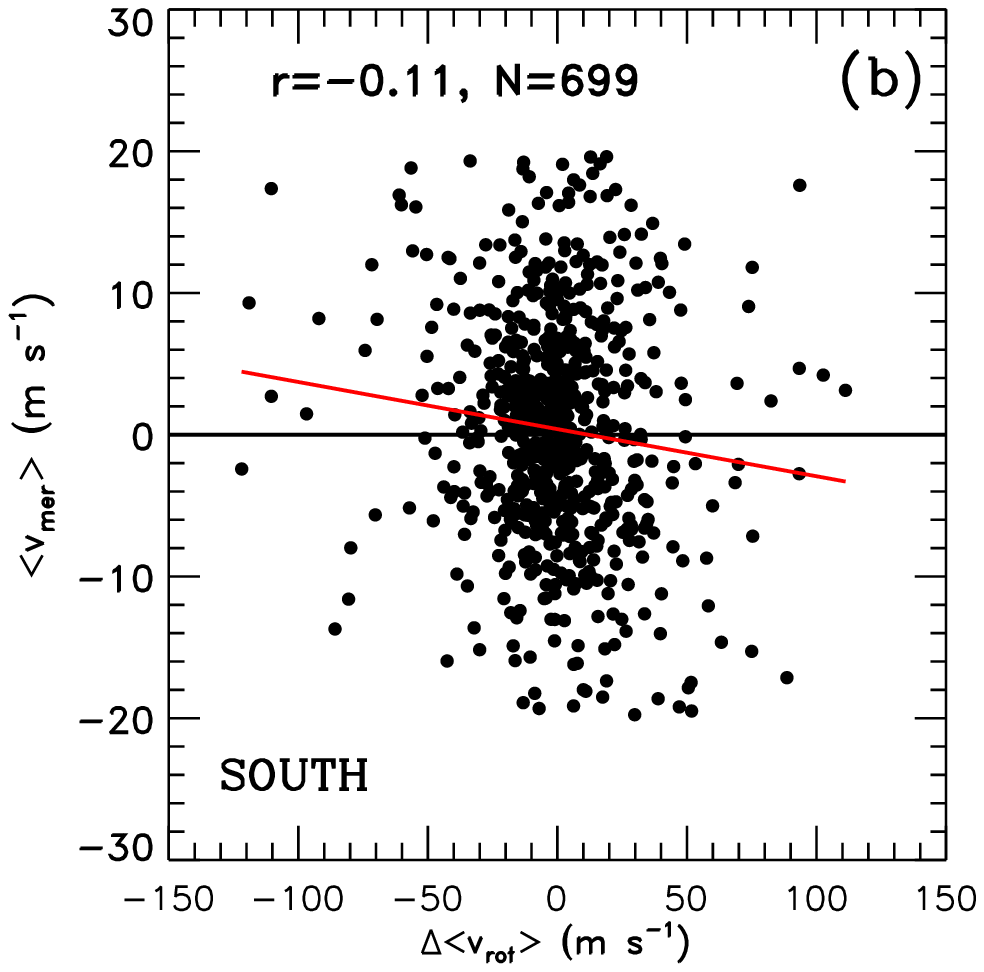}
\caption{Scatter plot of meridional velocity
($\langle v_{mer} (\theta) \rangle$ from $-20$ to $+20$ m s$^{-1}$)
 versus residual rotation rate
 ($\Delta \langle v_{rot} (\theta) \rangle$  $\approx -140$
 to $+140$ m s$^{-1}$) 
 determined by the combined data of sunspot groups  
({\bf a}) in northern hemisphere and ({\bf b}) in southern 
hemisphere  during the whole period 1874\,--\,2017.
The correlation coefficient  ($r$) and number of data points $N$,
is also shown.
Positive values  of meridional velocity indicate poleward motions and negative
 values indicate equatorward motions, both hemispheres.
The  continuous line (red) represents the linear best-fit.}.
\label{f1}
\end{figure}

\begin{table}
{\tiny
\caption[]{Values of the intercept ($C$)  and  slope ($D$),  and the 
corresponding  standard deviations  ($\sigma_C$)  and ($\sigma_D$) 
 of the linear-relationship between $\langle v_{mer} (\theta) \rangle$ and
$\Delta\langle v_{rot} (\theta) \rangle$ (equation~\ref{eq.1}) determined from the 
 northern  and southern hemisphere's
 sunspot-group data during  each  solar cycle 
($^{\mathrm a}$ indicates incomplete data) and
 during the whole period 1874\,--\,2017.
 The corresponding
values of the correlation coefficient ($r$), Students' t ($\tau$)
and probability ($P$) are given.
The $n$ represents  the number of data 
points in a solar cycle.}
\begin{tabular}{lccccccccc}
\hline
Cycle&Time& $C$&$\sigma_C$ & $D$& $\sigma_D$& $r$ & $\tau$& $P$&  $n$\\
\hline
&&&\multicolumn{4}{c}{Derived from northern hemisphere's data}\\
12&  1878--1889& $ 2.122$&   1.169& $ 0.028$&   0.036& $ 0.12$&   0.79&  0.784&  48\\
13&  1890--1901& $-2.346$&   1.195& $-0.030$&   0.035& $-0.12$&   0.85&  0.801&  51\\
14&  1902--1912& $ 0.079$&   1.374& $ 0.027$&   0.043& $ 0.10$&   0.62&  0.731&  44\\
15&  1913--1922& $-0.992$&   1.083& $-0.044$&   0.044& $-0.15$&   1.00&  0.839&  47\\
16&  1923--1932& $-0.318$&   0.903& $-0.024$&   0.027& $-0.12$&   0.89&  0.811&  52\\
17&  1933--1943& $-1.643$&   0.987& $-0.061$&   0.036& $-0.22$&   1.68&  0.951&  56\\
18&  1944--1953& $-2.517$&   0.909& $-0.092$&   0.041& $-0.31$&   2.28&  0.987&  52\\
19&  1954--1963& $-0.606$&   0.929& $ 0.005$&   0.033& $ 0.02$&   0.15&  0.559&  63\\
20&  1964--1975& $ 1.155$&   0.779& $-0.089$&   0.029& $-0.35$&   3.02&  0.998&  69\\
21&  1976--1985& $-0.236$&   1.156& $ 0.090$&   0.053& $ 0.23$&   1.71&  0.953&  56\\
22&  1986--1995& $ 1.641$&   0.942& $-0.057$&   0.039& $-0.20$&   1.45&  0.924&  54\\
23&  1996--2007& $ 2.280$&   0.973& $ 0.017$&   0.040& $ 0.05$&   0.43&  0.665&  65\\
24$^\mathrm{a}$&  2008--2017& $-1.839$&  1.374&  $-0.096$&   0.057& $-0.25$&   1.69&  0.951&  44\\
1874--2017&&     $-0.117$&  0.283&  $-0.020$&   0.010& $-0.07$&   2.00&  0.977& 711\\
\\
&&&\multicolumn{4}{c}{Derived from southern hemisphere's data}\\
12&  1878--1889& $ -0.355$&   1.079& $ -0.088$&   0.038& $-0.31$&   2.29&  0.987&  51\\
13&  1890--1901& $  1.281$&   1.190& $  0.010$&   0.046& $ 0.03$&   0.21&  0.585&  54\\
14&  1902--1912& $  0.068$&   1.080& $  0.032$&   0.038& $ 0.13$&   0.85&  0.800&  41\\
15&  1913--1922& $ -1.363$&   1.026& $ -0.127$&   0.044& $-0.41$&   2.90&  0.997&  44\\
16&  1923--1932& $ -0.444$&   1.163& $ -0.028$&   0.045& $-0.09$&   0.61&  0.729&  49\\
17&  1933--1943& $  0.527$&   0.909& $  0.076$&   0.036& $ 0.28$&   2.11&  0.980&  53\\
18&  1944--1953& $  0.242$&   1.081& $ -0.094$&   0.040& $-0.29$&   2.32&  0.988&  59\\
19&  1954--1963& $  1.039$&   1.008& $ -0.096$&   0.032& $-0.38$&   2.98&  0.998&  54\\
20&  1964--1975& $  0.842$&   1.077& $ -0.007$&   0.036& $-0.03$&   0.20&  0.581&  56\\
21&  1976--1985& $ -0.214$&   1.091& $ -0.022$&   0.039& $-0.07$&   0.56&  0.712&  61\\
22&  1986--1995& $  0.213$&   1.096& $ -0.041$&   0.045& $-0.12$&   0.92&  0.818&  57\\
23&  1996--2007& $  0.701$&   0.917& $ -0.054$&   0.045& $-0.15$&   1.19&  0.880&  66\\
24$^\mathrm{a}$&  2008--2017& $  3.145$&   1.221& $  0.019$&   0.061& $ 0.05$&   0.31&  0.620&  42\\
1874--2017&&     $  0.395$&   0.295& $ -0.033$&   0.011& $-0.11$&   2.98&  0.998& 699\\
\hline
\end{tabular}
\label{table1}
}
\end{table}

\begin{table}
{\tiny
\caption{Values of the intercept ($c$) and  the slope ($m$),  and the
corresponding standard deviations $\sigma_c$ and $\sigma_m$, respectively,
 determined  from the linear-least-square 
fits of the  DPD tilt angle data of Solar Cycles (SC)   21\,--\,23  to the 
equation~\ref{eq.2}.  The values of correlation coefficient ($r$),
$\chi^2$ and the corresponding probability ($P$), the ratio $m/\sigma_m$,
and the rms (root-mean-square deviation)  are also given.} 
\begin{tabular}{lccccccccc}
\hline
  \noalign{\smallskip}
SC&$c$&$\sigma_c$& $m$ &$\sigma_m$&$r$&$\chi^2$&
$P$&$m/\sigma_m$&rms\\
  \noalign{\smallskip}
\hline
&&&\multicolumn{4}{c}{From sunspot groups with whole-spot arreas}\\
&&&\multicolumn{4}{c}{Whole sphere}\\
 21&  1.88&  0.80&  0.26&  0.04&  0.79&  8.65&  0.19&  6.50&  1.94\\
 22&  0.58&  0.98&  0.36&  0.05&  0.97&  4.73&  0.58&  7.20&  1.05\\
 23&  1.50&  0.89&  0.36&  0.04&  0.97&  4.03&  0.67&  9.00&  1.13\\
&&&\multicolumn{4}{c}{Northern hemisphere}\\
 21&  2.40&  0.93&  0.25&  0.04&  0.79&  9.12&  0.17&  6.25&  1.86\\
 22& -0.73&  1.31&  0.42&  0.06&  0.93& 10.23&  0.12&  7.00&  1.89\\
 23&  1.66&  1.07&  0.36&  0.05&  0.75& 10.58&  0.10&  7.20&  2.85\\
&&& \multicolumn{4}{c}{Southern hemisphere}\\
 21&  1.06&  1.15&  0.29&  0.05&  0.62& 10.60&  0.10&  5.80&  3.05\\
 22&  1.30&  1.07&  0.34&  0.05&  0.91&  7.54&  0.27&  6.80&  1.77\\
 23&  0.82&  1.11&  0.41&  0.05&  0.97&  4.10&  0.66&  8.20&  1.09\\
\\
&&& \multicolumn{4}{c}{From sunspot groups with only umbrae areas}\\
&&& \multicolumn{4}{c}{Whole sphere}\\
 21&  2.04&  0.89&  0.28&  0.04&  0.96&  1.63&  0.95&  7.00&  0.93\\
 22& -1.72&  1.18&  0.48&  0.05&  0.99&  3.66&  0.72&  9.60&  0.92\\
 23&  0.20&  0.99&  0.39&  0.04&  0.90&  8.83&  0.18&  9.75&  1.98\\
&&& \multicolumn{4}{c}{Northern hemisphere}\\
 21&  1.98&  1.10&  0.31&  0.05&  0.87&  7.71&  0.26&  6.20&  1.84\\
 22& -2.47&  1.52&  0.53&  0.07&  0.88&  6.39&  0.38&  7.57&  3.04\\
 23&  0.16&  1.29&  0.42&  0.06&  0.68& 13.88&  0.03&  7.00&  3.93\\
&&& \multicolumn{4}{c}{Southern hemisphere}\\
 21&  1.75&  1.15&  0.27&  0.05&  0.93&  2.63&  0.85&  5.40&  1.20\\
 22& -1.17&  1.26&  0.44&  0.06&  0.94&  8.51&  0.20&  7.33&  1.92\\
 23& -0.11&  1.11&  0.39&  0.05&  0.97&  4.54&  0.60&  7.80&  1.02\\
\hline
  \noalign{\smallskip}
\end{tabular}
\label{table2}
}
\end{table}

\section{Results}
 Fig.~\ref{f1} shows  
the relationship between  meridional velocity
 and  residual rotation rate
 determined by the combined data of sunspot groups
 in northern  and  southern
hemispheres during the whole period 1874\,--\,2017.
In Table~\ref{table1} we have given the values   of the
 coefficients $C$ and $D$,
  and their uncertainties (values of standard deviation $\sigma$), 
of equation~\ref{eq.1},
i.e. the linear relationship between $\langle v_{mer} (\theta) \rangle$ and
$\Delta\langle v_{rot} (\theta) \rangle$  derived here from the 
 northern  and southern
hemispheres' sunspot group data  during each solar cycle and also 
from the data during the whole period 1874\,--\,2017.
 In the same table we have also 
given the  values of the  correlation coefficient ($r$), 
 Student's t ($\tau$) and the corresponding probability ($P$), 
minimum epoch of solar cycle, and number of data points. 
 These values indicate that     
the relationship between the meridional velocity and 
residual rotation rate  is reasonably good 
in some solar cycles and  is not good in some other solar cycles.
The value of $D$  determined from the whole period sunspot group
 data  is statistically significant on $2\sigma$ and  $3\sigma$ levels  
in the northern and southern hemispheres, respectively.
There exists a  north-south difference in the slope ($D$) but it 
is small (only at $1\sigma$-level). 
 Note that 
the values of $C$, $D$, etc. derived from the whole sphere sunspot-group 
 data were given in Table~2 of \cite{jj21}.

\begin{figure}
\centering
\includegraphics[width=8.5cm]{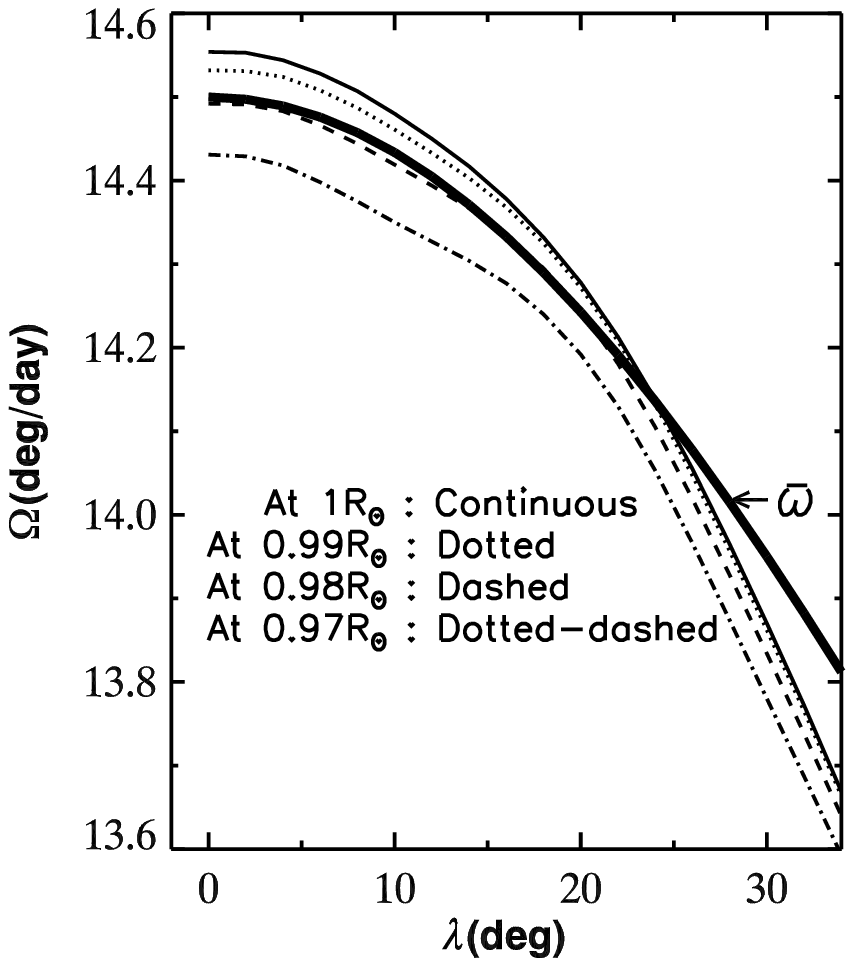}
\caption{Latitude ($\lambda$)  variations of the Sun's mean (over the 
whole period 1995\,--\,2009) internal
 rotation rate ($\Omega$) 
 at 0.97$R_\odot$, 0.98$R_\odot$, 0.99$R_\odot$, and 1$R_\odot$ 
(where $R_\odot$ is the radius of the Sun) 
derived by H. M. Antia from GONG data. {\bf {\it The thick-continuous 
curve represents}}  
 the latitude variation of the mean  angular velocity
 (${\bar \omega}$) obtained by using the grand differential rotation law 
derived from the sunspot-group data (cf. Sec. 2).}    
\label{f2}
\end{figure}

 Heliosesimic studies indicated that a net longitude-averaged
meridional flow converging onto the active latitudes 
\citep[][and references therein]{zhao04,hana22}. 
 Combined with the known torsional oscillation pattern
this yields a *positive* (i.e. poleward) net residual correlation in
plasma motions in the activity belt, opposite to that is found here
(overall D has negative sign). However, the measured meridioanal
 velocities have considerably large uncertainties. Already mentioned above  
 sunspots exhibit both equatoward and poleward meridional motions
\citep{ju06}. Generally, the studies based on
magnetic feature tracking~\citep[e.g][]{snod96}  suggest
equatorward angular momentum transport.

It has been argued that motions of magnetic tracers,
 such as  sunspots, are no tracers
of plasma motion. Instead, their motion 
is determined by a complex interplay
of magnetic forces and geometrical projection factors like the asymmetry of
emerging flux loops \citep{petrovay10}.   
 Difficult to  disprove this argument, and it is beyond the
scope of the present analysis. On the other hand, although 
sunspot motions may not be fully proxy
 of surface plasma motions,  to some extent proxy of the subsurface plasma
 motions \citep{howard96b}. 
  Fig.~\ref{f2} shows
latitude   variations of the Sun's mean (over the
whole period 1995\,--\,2009) internal
plasma rotation rates  
 at 0.97$R_\odot$, 0.98$R_\odot$, 0.99$R_\odot$, and 1$R_\odot$
(where $R_\odot$ is the radius of the Sun)
derived by H. M. Antia from the Global Oscillation Network Group (GONG) data.
In the same figure we have also shown  the latitude variation of the
 angular velocity (${\bar \omega}$) obtained by using the grand
 differential rotation law derived above from the sunspot-group data 
(see Sec. 2).
As we can see in this figure the differential rotational profile determined
from the sunspot-group data closely match with the corresponding
profile of the internal rotation rate at 0.98$R_\odot$ ($r = 0.999$),
 at least up to  about 25$^\circ$ latitude \citep[also see][]{jk99,jj13}.   
 Therefore, one can  believe
 that  the motions of sunspots 
are at least partially the tracers of plasma motions.
(In \cite{jj21} and here large restrictions were applied to the data. 
Consequently, we have obtained slightly more flat $\bar \omega( \lambda)$
 profile.)

Area-weighted heliographic positions of sunspot groups seem to be 
to a large extent increase uncertainties in the measured  velocities 
 \citep[e.g.][]{kut21}.  In the case of Greenwich sunspot data that
 we have used the positions of sunspot groups are  geometrical 
(not the area-weighted) positions of centers of groups. 
 The midpoints of sunspot groups  measured on  photographic plates were
 taken as positions of the groups \citep{poljancic11}. 

Hereafter the $D$ and $m$ that are derived from the data of the whole sphere, 
northern hemisphere, and southern hemisphere are indicated with 
suffixes {\rm W}, {\rm N}, and {\rm S}, respectively. 
In the case of the whole period, the north--south difference
 (north--south asymmetry), 
$D_{asym} = D_{\rm N} -D_{\rm S}$, is statistically significant. 
That is,  the uncertainty 
$\sigma_{asym}=\sqrt{(\sigma_{\rm N}^2/n_{\rm N}+\sigma_S^2/n_{\rm S})}$
  in $D_{asym}$ is  small  
 ($D_{asym} =0.013$ is about  three times larger than its 
uncertainty 0.004), where   
  $\sigma_{\rm N}$  and $\sigma_{\rm S}$ are the  values of the uncertainties 
($\sigma$ values) in $D_{\rm N}$  and $D_{\rm S}$, respectively,
and $n_{\rm N}$  and $n_{\rm S}$ are the number of velocity values
(values of $n$)  gone in the determinations of $D_{\rm N}$  and
 $D_{\rm S}$, respectively (see Table~\ref{table1}).
 The values of  $m_{\rm N}$ and $m_{\rm S}$  determined from the 
MWOB whole period 
1917\,--\,1986 tilt-angle data were given in Table~2 of \cite{jj23}.
The corresponding value of the north--south asymmetry, 
$m_{asym} = m_{\rm N} -m_{\rm S}$,  
is  statistically significant on 95\,\%  confidence level,
 i.e. $m_{asym} = 0.05$ is about 
 two times larger than the value  0.024 of its uncertainty
(the  value of $m_{asym}$ derived from the area-weighted tilt angle data
 is found  to be about five times larger than its uncertainty.)

\begin{figure}
\centering
\includegraphics[width=8.5cm]{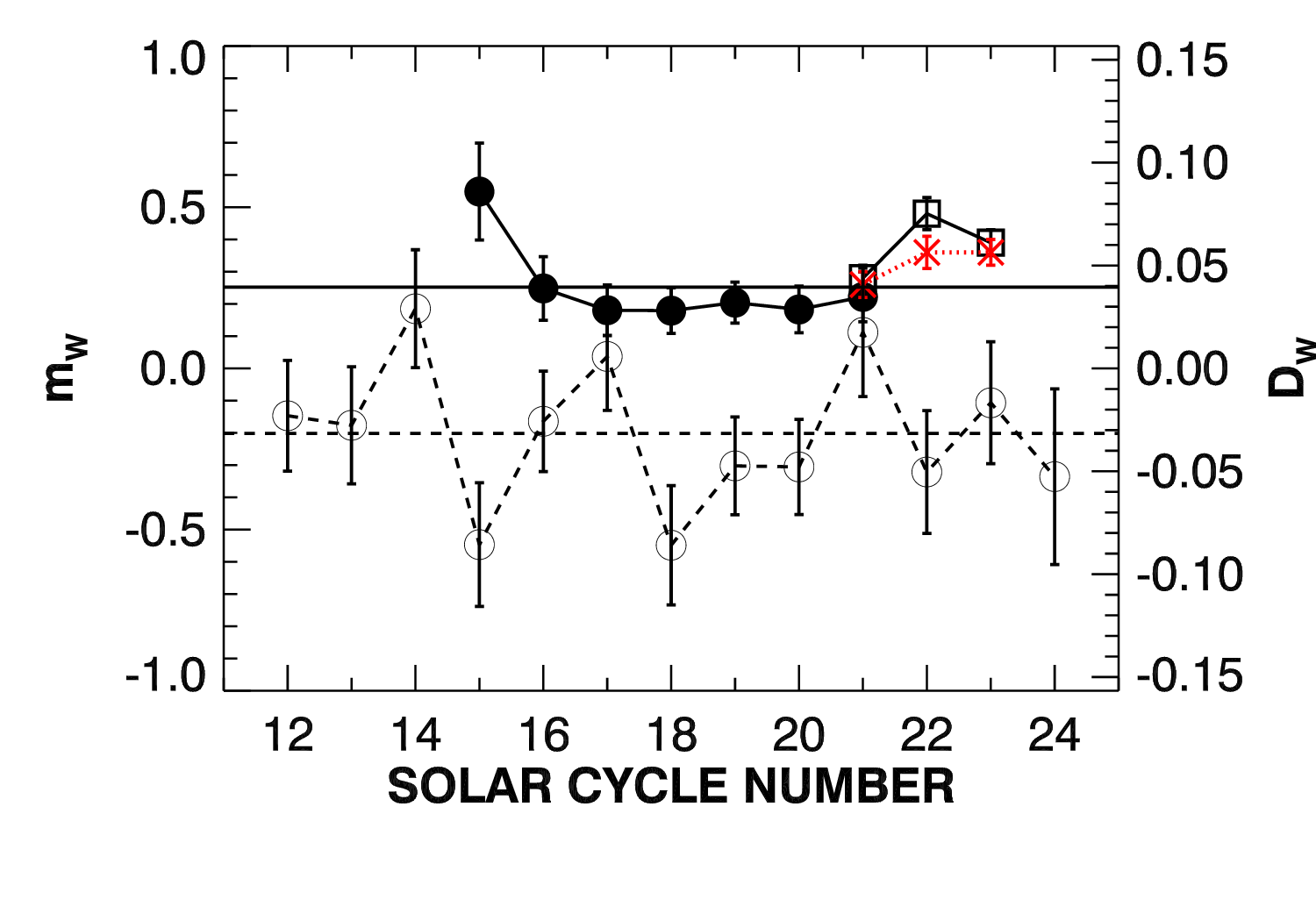}
\caption{Solar cycle-to-cycle variations in the coefficient of Joy's law 
($m_{\rm W}$, {\it filled circle-continuous curve})
derived from the MWOB whole-sphere sunspot-group data during Solar Cycles
15\,--\,21 and  the residual covariance  ($D_{\rm W}$,
 {\it open circle-dashed curve})
 derived from the combined GPR and DPD whole-sphere's sunspot-group data during 
Solar Cycles 12\,--\,24 (the MWOB data of Solar Cycle~15 and the DPD data of 
Solar Cycle~24 are incomplete). 
The mean values of the $m_{\rm W}$ and $D_{\rm W}$
 are shown by horizontal {\it continuous} and {\it dashed lines}, respectively.
The values of $m_{\rm W}$ (given in Table~\ref{table2})
 determined from the DPD tilt-angle data corresponding to the whole-spot areas 
({\it red cross-dotted curve}) and that corresponding to the umbrae areas
({\it square-continuous curve}) of sunspot groups are also shown.}
\label{f3}
\end{figure}

\begin{figure}
\centering
\includegraphics[width=8.5cm]{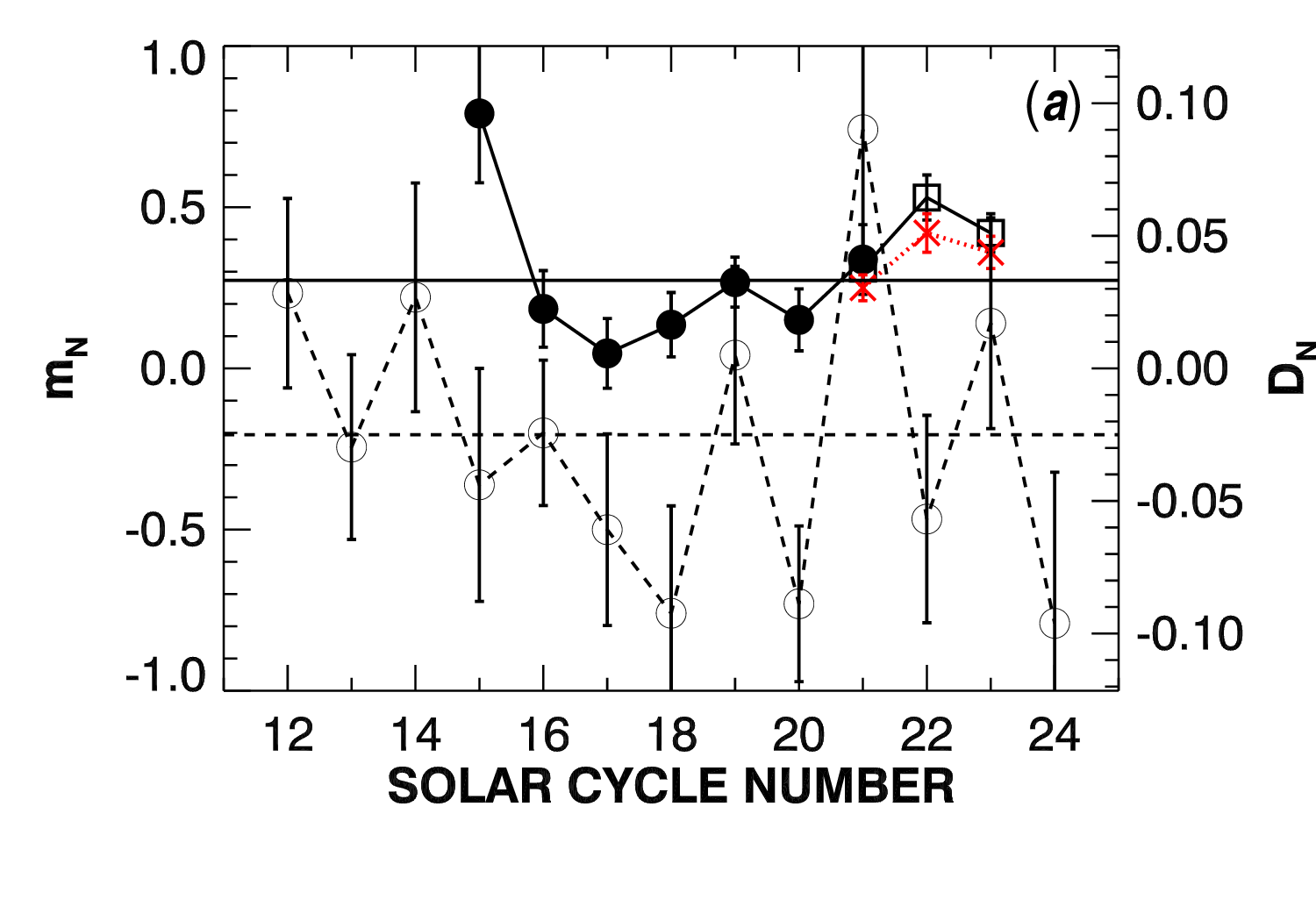}
\includegraphics[width=8.5cm]{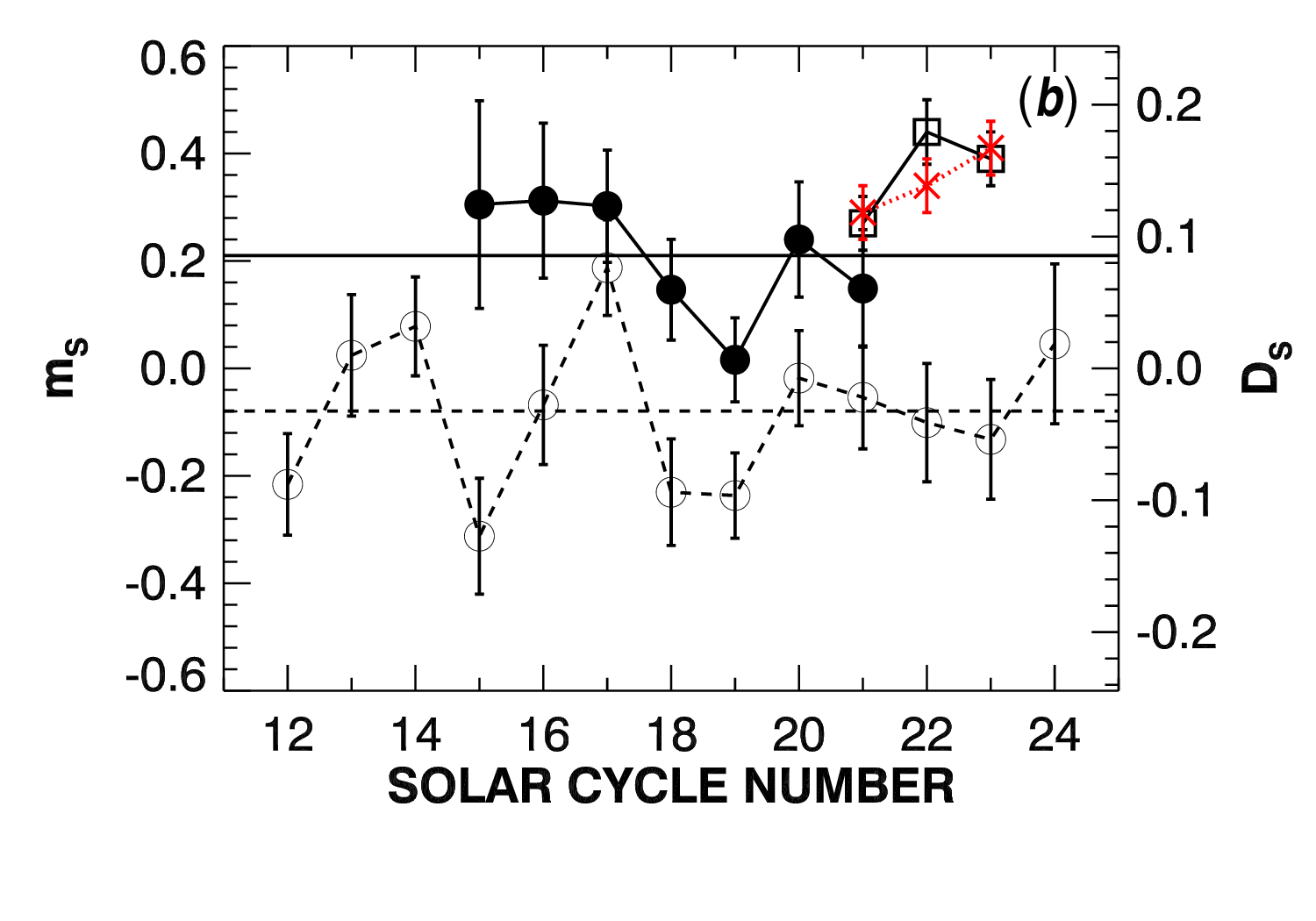}
\caption{Solar cycle-to-cycle variations in the coefficients of Joy's law
($m_{\rm N}$ and $m_{\rm S}$, {\it filled circle-continuous curves})
determined from the MWOB sunspot-group data during 
Solar Cycles 15\,--\,21 and the residual covariance
($D_{\rm N}$ and $D_{\rm S}$, 
{\it open circle-dashed curves}) determined from the GPR and DPD 
sunspot-group data during 
Solar Cycles 12\,--\,24, in ({\bf a}) northern and 
({\bf b})  southern hemispheres.  
The mean values of the $m_{\rm N}$ and $m_{\rm S}$
 are shown by horizontal {\it continuous lines}, and the mean values of the 
$D_{\rm N}$ and $D_{\rm S}$ are shown by horizontal {\it dashed lines}.
The values of $m_{\rm N}$ and $m_{\rm S}$ 
(given in Table~\ref{table2})
 determined from the DPD tilt-angle data corresponding to the whole-spot areas 
({\it red cross-dotted curve}) and that corresponding to the umbrae areas
({\it square-continuous curve}) of sunspot groups are also shown.}
\label{f4}
\end{figure}

\begin{figure}
\centering
\includegraphics[width=8.5cm]{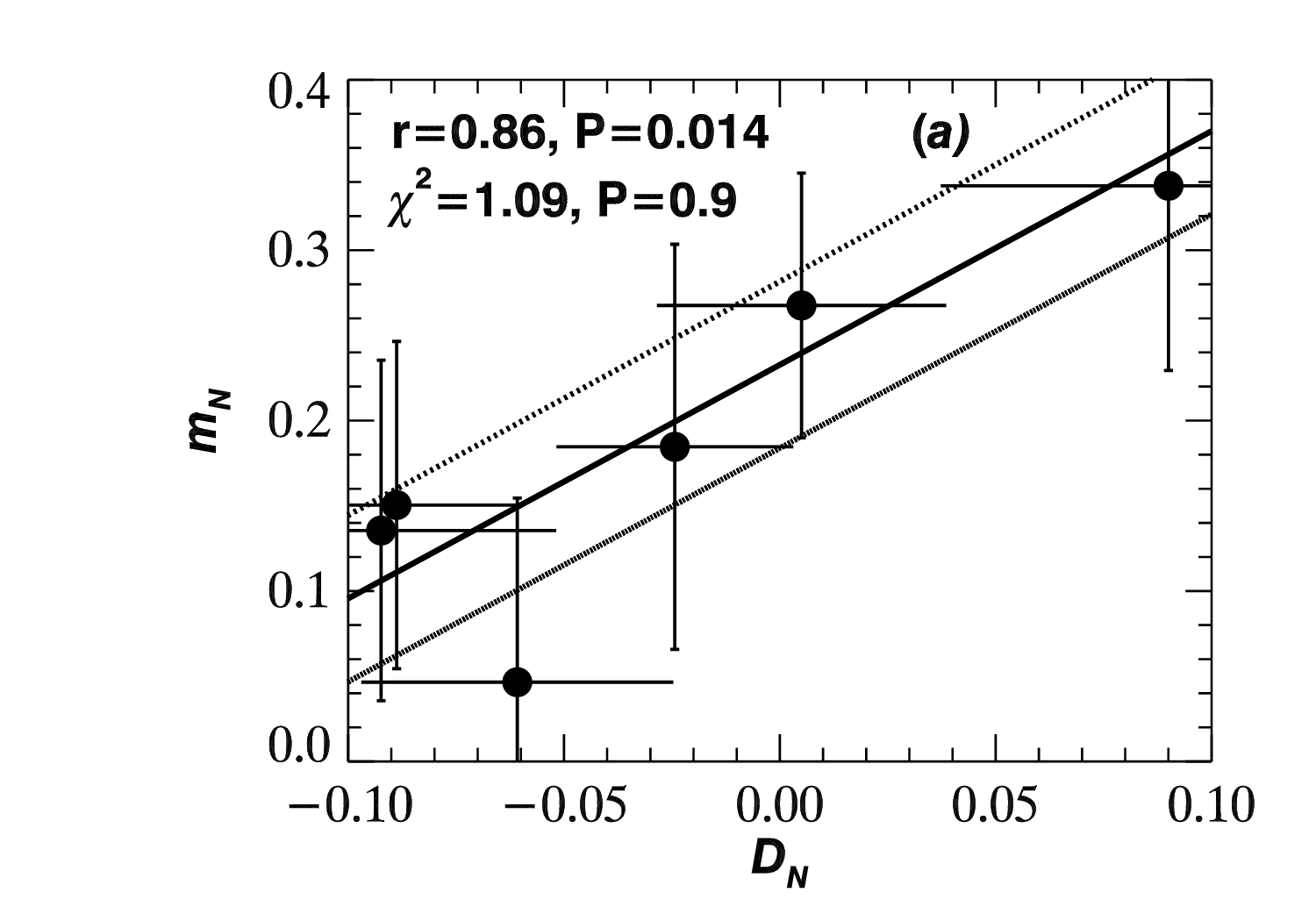}
\includegraphics[width=8.5cm]{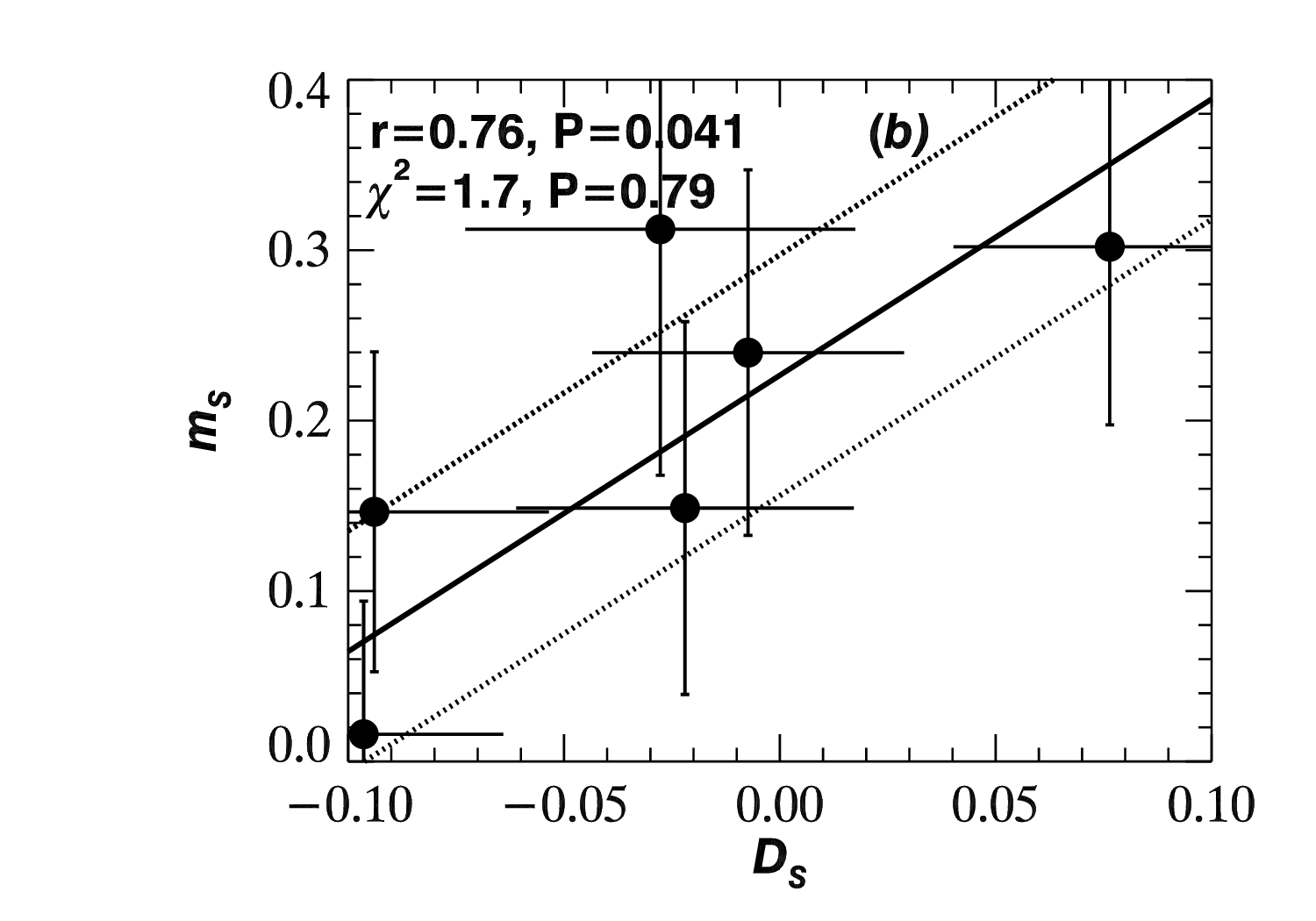}
\caption{Scatter plots of  residual covariance
 ($D_{\rm N}$ and $D_{\rm S}$)  verses the coefficient of  
Joy's law ($m_{\rm N}$ and  $m_{\rm S}$) 
derived from ({\bf a}) the northern hemisphere 
and ({\bf b}) the southern hemisphere sunspot-group data during 
Solar Cycles~16\,--\,21. The {\it continuous line} represents the
 best-fit linear relation and the {\it dotted lines} are drawn at 
one-rms ({\it root-mean-square deviation}) levels. The value of 
correlation coefficient ($r$) and $\chi^2$ 
and the values of the corresponding probabilities are also shown.} 
\label{f5}
\end{figure}

\begin{figure}
\centering
\includegraphics[width=8.5cm]{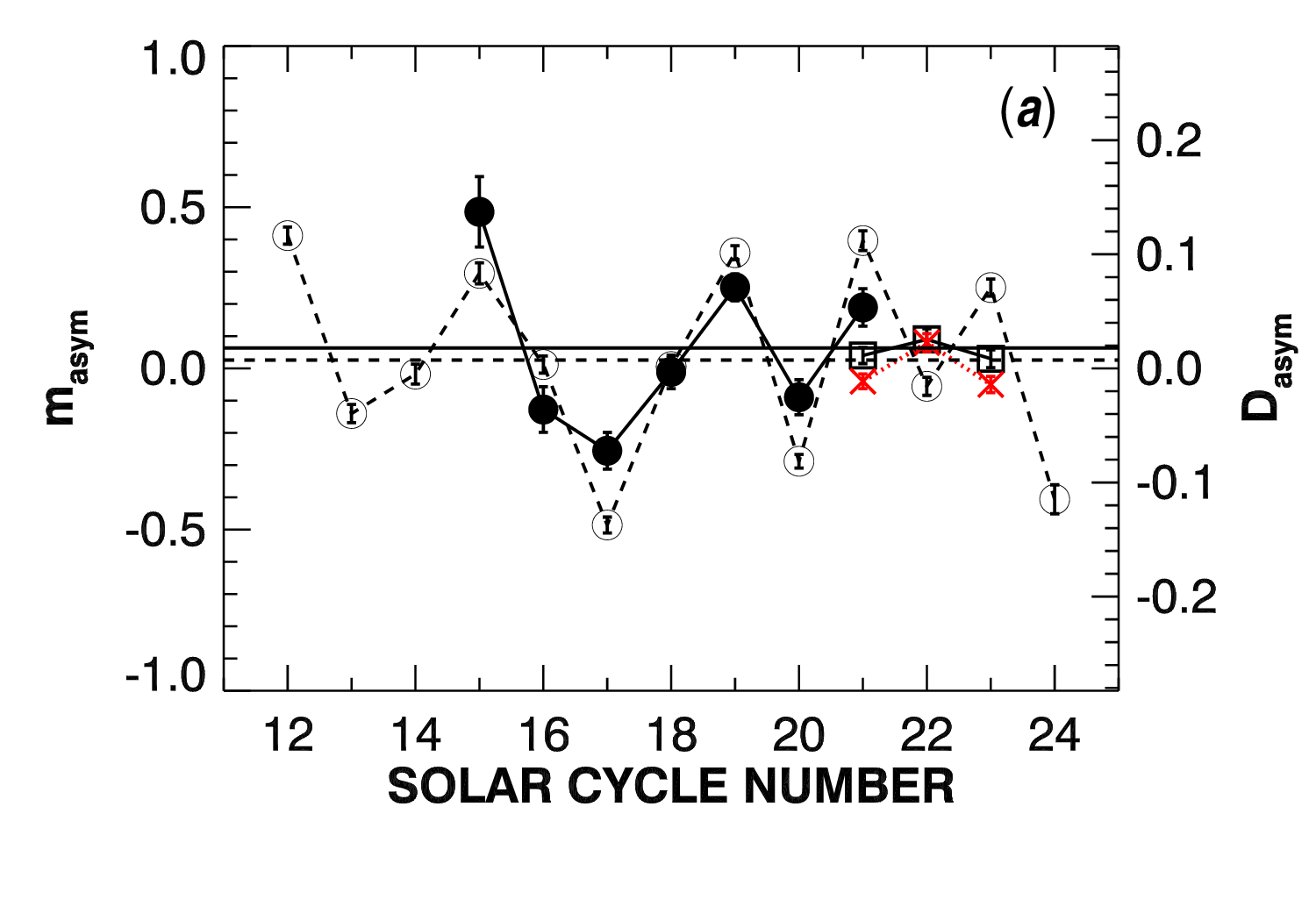}
\includegraphics[width=8.5cm]{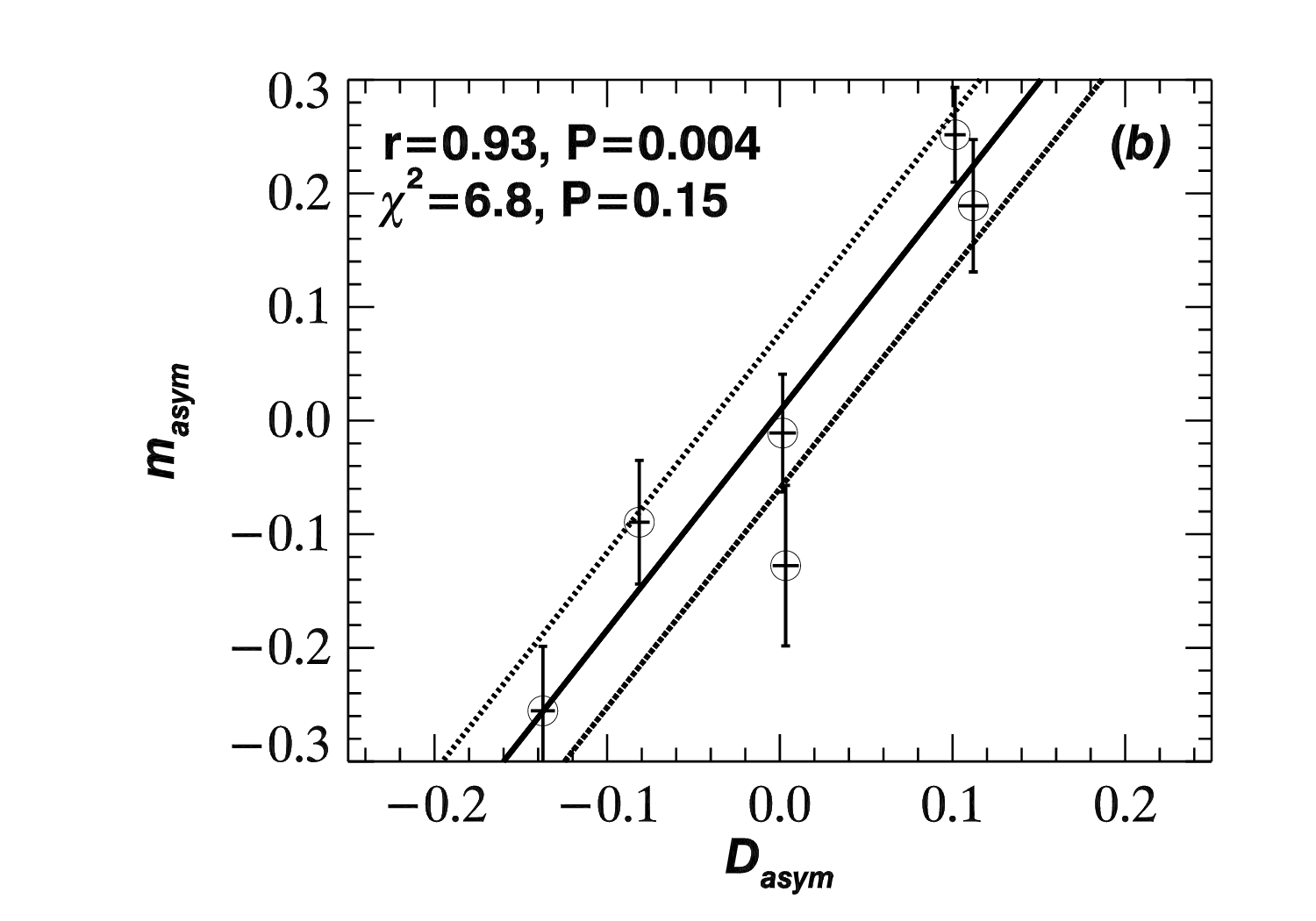}
\caption{({\bf a}) Solar cycle-to-cycle variations in
 north--south difference (north--south asymmetry) in  the
 coefficients of Joy's law ($m_{\rm asym}$: $m_{\rm N} - m_{\rm S}$,
 {\it filled circle-continuous curve}) determined from the MWOB sunspot 
group data during Solar Cycles 15\,--\,21 
and of the north--south asymmetry in  residual covariance 
 ($D_{\rm asym}$: $D_{\rm N} - D_{\rm S}$, 
{\it open circle-dashed curve}) derived from the combined GPR and DPD 
 sunspot-group data during Solar Cycles 12\,--\,24, in the northern 
and southern hemispheres. The horizontal {\it continuous} 
and {\it dashed lines} are drawn at the mean values of  $m_{\rm asym}$ 
and $D_{\rm asym}$,
respectively (note that the MWOB data of Solar Cycle~15 and the DPD data of 
Solar Cycle~24 are incomplete).
The values of $m_{\rm asym}$ 
 determined from the DPD tilt angles  corresponding to the whole-spot areas 
({\it red cross-dotted curve}) and that corresponding to the umbrae areas
({\it square-continuous curve}) of sunspot groups are also shown.
({\bf b}) Scatter plot of  $D_{\rm asym}$ versus $m_{\rm asym}$ 
during Solar Cycles 16\,--\,21. 
  The {\it continuous line} represents the best fit linear
relation and the horizontal {\it dotted lines} are drawn at one-rms levels. 
The values of correlation coefficient ($r$) and $\chi^2$
and the corresponding probabilities are also shown.
} 
\label{f6}
\end{figure}

In Table~\ref{table2} we have given the values of the parameters of
 the Jay's laws, i.e.  
the linear relationships of tilt angles  and the   latitudes  of sunspot 
groups,  determined from the DPD tilt-angle data
corresponding to the whole-spot areas and only the umbrae areas 
of the sunspot groups in the  whole sphere, and the northern and the 
southern hemispheres
during  each of the complete Solar Cycles 21\,--\,23. In most of the cases  
 the linear relationship is reasonably good, i.e. the
 correlation is statistically significant, $\chi^2$ is insignificant 
(the 95\% significant value of $\chi^2$ is 14.067 for 7 degrees of freedom),
 and the ratio of the slope to  its uncertainty is reasonably high.
In each of the three cases: whole sphere, northern hemisphere,
 and southern hemispheres,
the pattern of the slope of Joy's law during Solar Cycles 21\,--\,23 
determined from the DPD tilt-angle data  is slightly different from that 
obtained by \cite{jiao21} using the DPD area-weighted tilt angles and
 forcing the linear function pass through the origin.   
     
Fig.~\ref{f3} shows the solar cycle-to-cycle variations in $m_{\rm W}$
 derived from the MWOB whole-sphere's sunspot-group data during Solar Cycles
 15\,--\,21 \citep[the values are taken from Table~3 of][]{jj23}  and
 $D_{\rm W}$ \citep[the values are taken from Table~2 of][]{jj21} 
   derived from the combined GPR and DPD whole-sphere's sunspot-group data
  during Solar Cycles 12\,--\,24 (note that the MWOB data of Solar Cycle~15 
and the DPD data of Solar Cycle~24 are incomplete). In the same figure we have
 also shown the values of $m$   determined here
 from the DPD tilt-angle data of Solar Cycles 21\,--\,23,  for both the cases
 of whole-spot areas and only umbrae areas of sunspot groups.
 As we can see in this figure only the value of 
$m_{\rm W}$ of the incomplete Solar Cycle~15 significantly differs 
(large) the corresponding over all cycles' mean value (the mean value of
 $m_{\rm W}$ is calculated from the MWOB data only). The $m_{\rm W}$ is almost
 constant during Solar Cycles 16\,--\,21. The values of $m_{\rm W}$  
(given in Table~\ref{table2} above) determined from the DPD data of 
 tilt angles
corresponding to both the  whole-spot areas and only umbrae areas  
of sunspot groups are slightly larger than the mean value.
 In Solar Cycles~14, 17, and 21,  
the values of  $D_{\rm W}$ are statistically significant and have a positive
 sign,  whereas the corresponding values in Solar Cycles~15 and 18 are
 significant, but have a negative sign. It seems there exist  3\,--\,5
 solar-cycle variations in $D_{\rm W}$. Obviously, there exists no significant 
correlation/anticorrelation 
 between  $m_{\rm W}$ and $D_{\rm W}$ during Solar Cycles~15\,--\,21
 ($r = -0.44$ only). 
In the case of Solar Cycles~21\,--\,23 the DPD data indicate an
 anti-correlation between $m_{\rm W}$  and $D_{\rm W}$.
The overall cycles' mean $m_{\rm W}$ is positive because in  a large number
 of solar cycles 
$m_{\rm W}$ is positive, i.e. the leading portions of a large number of
the  sunspot groups  closer to the equator than their respective 
following portions, in   consistent with Joy's law. 
The overall cycles' mean $D_{\rm W}$ is negative means that in a large number 
of solar cycles the angular momentum transport is mostly toward equator.

Figs.~\ref{f4}a and \ref{f4}b show the solar cycle-to-cycle variations in
 the coefficients of Joy's law 
\citep[the values are taken from Tables~4 and 5 of][]{jj23}
 and of  
the coefficients of equatorward/poleward angular momentum transport 
in the northern and southern hemispheres (values given in
 Table~\ref{table1} above), respectively. As we can see 
in this figure the patterns of the variations in 
the  coefficients $m_{\rm N}$ and $m_{\rm S}$ 
of Joy's law determined from the MWOB data in the northern and 
southern hemispheres are considerably different. The patterns of the
 variations in the slopes $D_{\rm N}$ and $D_{\rm S}$, i.e. the coefficients
 of  the angular momentum transport  in the northern and southern hemispheres, 
respectively, are also  considerably different. The variation of $m_{\rm N}$
 is in a large extent the same as that of the $m_{\rm W}$ shown
 in Fig.~\ref{f3},
 but it looks to be correlate with the variation in $D_{\rm N}$ during 
Solar Cycles~16\,--\,21.
 The variation of $m_{\rm S}$  also looks to be correlate with $D_{\rm S}$ 
during Solar Cycles 16\,--\,21.
 However, both $m_{\rm N}$ and $m_{\rm S}$
 (values given in Table~\ref{table2} above) 
 determined from the DPD tilt-angle data look to be 
slightly anti-correlate with the corresponding $D_{\rm N}$ and  $D_{\rm S}$ 
during Solar Cycles 21\,--\,23.
 Both $m_{\rm N}$ and $m_{\rm S}$
  have positive values in most solar cycles (consistent with Joy's law), 
whereas both 
$D_{\rm N}$ and $D_{\rm S}$  have negative values in many 
solar cycles (equatorward angular momentum transport).
The long-term trends of the variations in  $m_{\rm N}$
 and $m_{\rm S}$  (also in $m_{\rm w}$, see Fig.~\ref{f3}) of MWOB and DPD data
together suggest  the existence of a long-term (around 8-solar
 cycle) cycle (Gleissberg cycle) in these parameters.

Since the first four years' MWOB data of Solar Cycle~15 are missing, 
the corresponding values of  $\langle \bar \gamma \rangle$ are not
 well determined and hence the slope of 
this solar cycle might be also not  well determined. 
Hence, we calculated the 
 correlation between the  coefficients of
Joy's law and  residual covariance with and without including the data 
points of this solar cycle.
 In fact,  when they  were included 
the correlation was found to be very low. Moreover, the data points of 
the $m_{\rm N}$ and $m_{\rm S}$ 
of Solar Cycles~21\,--\,23  determined from the DPD tilt-angle data 
were not considered here and not included in the determination  of 
correlation. This is because as we have already seen in Fig.~\ref{f4} above, 
  their behavior is different and  
seem to be anti-correlate with  the corresponding 
$D_{\rm N}$ and $D_{\rm S}$, respectively.
   Fig.~\ref{f5} shows the correlation between the  coefficients of
Joy's law and angular momentum transport in 
 the northern  and  southern hemispheres
 during Solar Cycles~16\,--\,21. The correlation is highly statistically 
significant (more than 95\,\% confidence level) in the 
northern hemisphere, and is about 95\,\% confidence level in 
the southern hemisphere.
 We obtained the following relations: 
\begin{equation}
m_{\rm N} = (1.37\pm0.82) D_{\rm N} + 0.22\pm0.05,   \ {\rm and} 
\label{eq.3}
\end{equation}
\begin{equation}
m_{\rm S} =(1.62\pm0.84) D_{\rm S} + 0.23\pm0.05.   
\label{eq.4}
\end{equation}
The values of $\chi^2$ of the  corresponding least-square best fits 
are small (statistically   insignificant).
The relationship (equation~\ref{eq.3}) in the northern
 hemisphere is slightly more accurately determined  than that
 (equation~\ref{eq.4}) 
 in the southern hemisphere (see Fig.~\ref{f5}). There is a suggestion that the
  variations of $m_{\rm S}$ with $D_{\rm S}$ are steeper than the variations of
 $m_{\rm N}$ with $D_{\rm N}$. 
Overall, we can conclude that there exists a reasonably significant 
linear relationship between the  coefficients of
Joy's law and angular momentum transport in both the hemispheres 
during Solar Cycles 16\,--\,21. 
There exists no significant correlation between $m_{\rm W}$ and $D_{\rm W}$. 
But there exists a reasonable correlation between the corresponding 
parameters of an hemisphere. 
This conveys that it is necessary and important to  determine 
  Joy's law and   residual covariance  and their relationship from the
hemispheres' data separately. 

Fig.~\ref{f6}a shows the  solar cycle-to-cycle variation in the 
 north--south difference, i.e. north--south asymmetry 
($m_{\rm asym}$: $m_{\rm N} - m_{\rm S}$), 
in  the coefficients of Joy's law
  determined from the MWOB sunspot-group data during Solar Cycles 15\,--\,21, 
and the  north--south asymmetry in  residual covariance
 ($D_{\rm asym}$: $D_{\rm N} - D_{\rm S}$) determined from the 
combined GPR and DPD sunspot-group data during Solar Cycles 12\,--\,24. 
The variation in $m_{\rm asym}$ determined from the DPD tilt-angle data during
 Solar Cycles~21\,--\,23 is also shown.
 As can be seen in this figure the cycle-to-cycle variations of the  
$m_{\rm asym}$ and $D_{\rm asym}$  are closely resemble each other
 during Solar Cycle 15\,--\,21, in such way that one can think that 
the behavior of  $m_{\rm asym}$ before Solar Cycle~15 and after
 Solar Cycle~21  might be also similar as that of $D_{\rm asym}$. 
But it seems  not, i.e.  the  $m_{\rm asym}$ determined from
 the DPD tilt-angle data is not 
significantly different from zero during Solar Cycles~21\,--\,23 and 
seems to be to some extent  anticorrelate to the corresponding $D_{\rm asym}$.  
We calculated the correlation between the 
$m_{\rm asym}$ and $D_{\rm asym}$  with and without the data points of 
Solar Cycle~15 (the values of $m_{\rm asym}$ determined from the DPD data
 are not considered). With including the data points of Solar Cycle~15 we 
obtained 
$r = 0.84$ and without including these data points 
 we obtained $r = 0.92$, both these are statistically  significant on more 
than 95\,\% confidence level. 
Fig.~\ref{f6}b shows  the  correlation between the $D_{\rm asym}$ and
 $m_{\rm asym}$
during Solar Cycles 16\,--\,21. We obtained the following linear 
relationship between the $m_{\rm asym}$ (derived from the MWOB data)
 and $D_{\rm asym}$ (derived from the GPR and DPD data) during Solar Cycles
 16\,--\,21:
\begin{equation}
m_{\rm asym} = (1.93 \pm 0.25) D_{\rm asym} + 0.008\pm0.022. 
\label{eq.5}
\end{equation}
The corresponding least-square best fit of this equation is very good, 
i.e. the value of the 
slope is about 7.7 times larger than the corresponding standard deviation and
 $\chi^2$ is reasonably small ($< $95\,\% significant level). In addition, 
 value of the intercept is negligibly small, suggesting that there exists  
 one-to-one correspondence between $m_{\rm asym}$ and $D_{\rm asym}$. 

In Figs.~\ref{f4} and \ref{f5}  
there is also a suggestion of  weak and  strong Joy's laws largely
 associated with equatorward and poleward angular momentum transport, 
respectively, 
largely implying that the strength of the Joy's law depends on the
 strength of the poleward angular momentum transport.
Overall, the results above, indicate  that there
exists a role of surface/subsurface poleward/equatorward angular
momentum transport in the Joy's law of tilt angles of bipolar active
regions (sunspot groups). Anti-Joy regions (those with the
follower portion closer to the equator than the leading portion) 
may be  mostly related to the equatorward angular momentum transport 
(return flow in relatively deeper layers). 

\section{Discussion and Conclusions}
  Coriolis force may be responsible for Joy's
law and hence the latter may be related to the equatorward/poleward
angular momentum transport. Here we find that  there exists around 8-solar
 cycle (Gleissberg cycle) trend in the long-term variation of the slope
of Joy's law (increase of tilt angle with latitude).
There exists a reasonably significant correlation
between the solar cycle modulations in the coefficients of 
Joy's law and residual covariance  determined  
from the northern and southern hemispheres' sunspot-group data  during
Solar Cycles~16\,--\,21.  There exists also
a good correlation between the north--south differences of these 
 parameters. We consider the residual covariance $D$ tentatively
  represents the coefficient of  angular momentum transport.
 Overall these results indicate that there 
exists a role of surface/subsurface poleward/equatorward angular
momentum transport (latitude dependent Reynolds tress)  in the Joy's law 
of tilt angles of bipolar active regions.  That is, there is a suggestion of 
 the strength of the Joy's law depends on the
 strength of the poleward angular momentum transport.

The existence of a relationship between the tilt angles of active regions 
and the rotation rates of the active regions have been found.
 \cite{howard96a} found 
that  sunspot groups with the tilt angles near the average value or 
$0^\circ$, rotate significantly faster ($\approx$1\,\% between tilt angles of 
$0^\circ$ and $45^\circ$) than do groups having tilt angles far from this 
value. There is a latitude dependence of this effect that follows 
Joy's law~\citep{howard96c}.  However, in the present analysis we find no
 significant correlation between the mean residual rotation rate or the mean
 meridional velocity  of the sunspot groups of a solar cycle  
and the coefficient of Joy's law of the solar cycle. The latter is found to
 be to  some extent correlate ($r = 0.78$) with magnitude of the 
coefficient of differential
 rotation (magnitude of the latitude gradient of rotation) derived from 
sunspot group data in the northern hemisphere only (not in the southern 
hemisphere), suggesting that in the northern hemisphere an increase in the 
strength of the Joy's law to some extent associated with an increase in
the strength of  differential rotation. 
 The aforementioned result that the existence of a reasonable 
correlation  
between the residual covariance (coefficient of angular 
momentum transport) and coefficient of Joy's law 
seems to be consistent with the conclusions of  the models that invoke the
 effect 
of  Coriolis force for cause of tilt angles of bipolar magnetic regions 
and  Joy's law~\citep[e.g.][]{dsilva93,fisher95}. 

Earlier, \citep{jj23}, we found that there exists a significant   
anti-correlation between $m_{\rm S}$ derived from the MWOB tilt-angle data
 and the  amplitude ($R_{\rm M}$) of a 
solar cycle and no a significant correlation exists between $m_{\rm N}$ 
and $R_{\rm M}$. We found 
the $m_{\rm asym}$ of a solar cycle determined from the area-weighted tilt 
angles  is significantly correlating  with the amplitude ($R_{\rm M}$) of 
the solar cycle and in the case of the tilt angles (not area-weighted), 
the correlation between $m_{\rm asym}$ 
 and $R_{\rm M}$ was found to be statistically insignificant.
It should be noted here that in the present analysis all the results are based 
on the tilt-angle data (no significant result is found 
from the area-weighted tilt-angle data). There exists no  significant 
correlation between $R_{\rm M}$ and  residual covariance 
 in all the three cases: whole sphere, northern hemisphere, 
and southern hemisphere. Therefore, the  effect of evolution
 of sunspot groups is less or absent in the present results. 

Since here we find in both the hemispheres the coefficient of
 Joy's law is correlating to the residual covariance, 
 hence we can conclude that  angular momentum transport  
has an important role in the mechanism behind the Joy's law.
 The strength of the Joy's law seems to be  depending on the
 strength of the poleward angular momentum transport.
 However, further studies are necessary to find out the significant
 implications of the present results and their connection to the 
cycle-to-cycle modulation in the amplitudes of solar cycles.

Earlier no correlation was found between the slope of Joy's law of 
a solar cycle  
determined from the tilt angles of sunspot groups measured in 
Kodaikanal Observatory (KOB)  and the amplitude of the solar cycle 
\citep[see][]{jj23}. Here also  
no significant correlation is found between the slope of Joy's law 
determined from the KOB  tilt-angle data 
 and the residual covariance. 
However, we have used the KOB sunspot data that are available at
the website {\sf www.ngdc.noaa.gov/stp/solar/sunspotregionsdata.html}.
The aforementioned results need to be verified by using the data of
 the digitized  white-light images of KOB \citep{rcj21}.  
\cite{tlat18} have studied  Joy's law by using the digitized sunspot drawings
 from Mount Wilson Observatory in Solar Cycles~15\,--\,24 and  found 
no significant correlation between the mean tilt angle and the amplitude of a
 solar cycle. Here we also 
determined the coefficients of Joy's law  from the mean values of the tilt 
angles in different $10^\circ$ latitude intervals (having $5^\circ$ overlaps)
 given in Table~2 of \cite{tlat18}  for each of Solar Cycles~15\,--\,24. 
In this case also no significant correlation is found between the  slope
 of Joy's law and the coefficient of angular momentum transport. This implies 
 there exists a kind of uniqueness in the MWOB tilt angle dataset that were used
 here compared to the right now available other tilt angle datasets.

\section*{Acknowledgements}
The author thanks the anonymous referee for useful comments and
suggestions.  He thanks  Manjunath Hegde for fruitful
 discussion.  He acknowledges the work of all the
 people who contribute to and maintain the GPR, DPD, 
MWOB and KOB  sunspot databases. He used the  maximum and
 minimum epochs  of Solar Cycles 12\,--\,24 determined by \citet{pesnell18}
from the time series of  13-month smoothed monthly mean
values of  version~2 of international sunspot number (SN) available
 at {\textsf www.sidc.be/silso/datafiles}.  The  Sun's internal
 rotation rates derived from the GONG measurements during 1995\,--\,2009 was
 provided by H. M. Antia.

\section*{Data Availability}

 All data generated or analysed during this study are included in this article.



\bibliographystyle{mnras}
\bibliography{joy} 


\bsp    
\label{lastpage}
\end{document}